\documentclass{amsart}
\usepackage{amssymb}
\usepackage{graphicx}
\usepackage{amscd}
\usepackage{amsmath}

\theoremstyle{plain}

\newtheorem{proposition}{Proposition}
\newtheorem{remark}{Remark}

\numberwithin{equation}{section}

\input{tcilatex}

\begin{document}
\title[Stochastics as a Dirac Boundary-value Problem.]{On Stochastic Schr%
\"{o}dinger Equation as a Dirac Boundary-value Problem, and an Inductive
Stochastic Limit. }
\author{V P Belavkin}
\address{School of Mathematics, Nottingham University, \\
NG7 2RD, UK}
\email{vpb@maths,nott.ac.uk}
\urladdr{www.maths.nott.ac.uk/personal/vpb/}
\thanks{ Published in: \textit{Evolution Equations and Their Appications},
Marcel Dekker, Inc: New York 2000, pp. 311--327.}
\thanks{This work was supported by Royal Society grant for UK-Japan
collaboration and EPSRC research grant GR/M66196.}
\date{January 12, 2000}
\dedicatory{Dedicated to Sergio Albeverio.}
\maketitle

\begin{abstract}
We prove that a single-jump quantum stochastic unitary evolution is
equivalent to a Dirac boundary value problem on the half line in one extra
dimension. It is shown that this exactly solvable model can be obtained from
a Schr\"{o}dinger boundary value problem for a positive relativistic
Hamiltonian in the half-line as the inductive ultrarelativistic limit,
correspondent to the input flow of Dirac particles with asymptotically
infinite momenta. Thus the problem of stochastic approximation is reduced to
the to the quantum-mechanical boundary value problem in the extra dimension.
The question of microscopic time reversibility is also studied for this
paper.
\end{abstract}

\section{Introduction.}

The stochastic evolution models in a Hilbert space have recently found
interesting applications in quantum measurement theory, see for example the
review paper \cite{AKS97}. In this paper we are going to show on a simple
example that classical discontinuous stochastics can be derived from a
quantum continuous deterministic conservative dynamics starting from a pure
quantum state. It has been already proved in \cite{Bel95} that the piecewise
continuous stochastic unitary evolution driven by a quantum Poisson process
is equivalent to a time-dependent singular Hamiltonian Schr\"{o}dinger
problem, and the continuous stochastic unitary evolution driven by a quantum
Wiener process can be obtained as the solution of this problem at a central
limit. Here we want to start form a non-singular time-independent dynamics.

There exits a broad literature on the stochastic limit in quantum physics in
which quantum stochastics is derived from a nonsingular interaction
representation of the Schr\"{o}dinger initial value problem for a quantum
field by rescaling the time and space as suggested in \cite{AAFL91}. Our
intention is rather different: instead of rescaling the interaction
potentials we treat the singular interactions rigorously as the boundary
conditions, and obtain the stochastic limit as an ultrarelativistic limit of
the corresponding Schr\"{o}dinger boundary value problem in a Hilbert space
of infinite number of particles. We shall prove that the discontinuous and
continuous quantum stochastic evolutions can be obtain in this way from a
physically meaningful time continuous (in strong sense) unitary evolution by
solving a boundary value problem with an initial pure state in the extended
Hilbert space.

First we shall describe the boundary value problem corresponding to the
single-point discontinuous stochastic evolution and demonstrate the
ultrarelativistic limit in this case. Then the piece-wise continuous
stochastic evolution and the continuous diffusive and quantum stochastic
evolution can be obtained as in \cite{Bel95, Bel96}. But before to perform
this program, let us describe the unitary toy model giving an ''unphysical''
solution of this problem corresponding to the free hamiltonian $h\left(
p\right) =-p$. This toy model in the second quantization framework was
suggested for the derivation of quantum time-continuous measurement process
in \cite{Bel94}. Recently Chebotarev \cite{Che97} has shown that the
secondary quantized time-continuous toy Hamiltonian model in Fock space with
a discontinuity condition is equivalent to the Hudson-Parthasarathy (HP)
quantum stochastic evolution model \cite{HP84} in the case of commuting
operator-valued coefficients of the HP-equation. Our approach is free from
the commutativity restriction for the coefficients, and we deal with
time-reversible Dirac Hamiltonian and the boundary rather than physically
meaningless discontinuity condition and time irreversible $-p$. Moreover, we
shall prove that the stochastic model can be obtained from a positive
relativistic Hamiltonian as an inductive ultra relativistic limit on a union
of Hardy class Hilbert spaces. We call this limit the inductive stochastic
approximation.

\section{A toy Hamiltonian model.}

Here we demonstrate on a toy model how the time-dependent single-point
stochastic Hamiltonian problem can be treated as an interaction
representation of a self-adjoint boundary-value Schr\"{o}dinger problem for
a strongly-continuous unitary group evolution.

Let $\mathcal{H}$ be a Hilbert space, $H$ be a bounded from below
selfadjoint operator, and $S$ be a unitary operator in $\mathcal{H}$, not
necessarily commuting with $H$. The operator $H$ called Hamiltonian, is the
generator for the conservative evolution of a quantum system, described by
the Schr\"{o}dinger equation $i\hbar\partial_{t}\eta=H\eta$, and the
operator $S$ called scattering, describes the unitary quantum jump $%
\eta\mapsto S\eta$ of the state vectors $\eta\in\mathcal{H}$ caused by a
singular potential interaction in the system, with the continuous unitary
evolution $\eta\mapsto e^{-\frac{i}{\hbar}tH}\eta$ when there is no jump. As
for an example of such jump we can refer to the von Neumann singular
Hamiltonian model for indirect instantaneous measurement of a quantum
particle position $x\in\mathbb{R}$ via the registration of an apparatus
pointer position $y\in\mathbb{R}$. It can be described \cite{Bel95, Bel96}
by the $x$-pointwise shift $S$ of $y$ as the multiplication $\hat{\sigma}$
by $\sigma\left( x\right) =e^{x\partial_{y}}$ in the Hilbert space $%
L^{2}\left( \mathbb{R}^{2}\right) $ of square-integrable functions $%
\eta\left( x,y\right) $, and it does not commute with the free Hamiltonian
operator $H=\frac{1}{2}\left( y^{2}-\partial_{x}^{2}\right) $ say, of the
system ''quantum particle plus apparatus pointer''.

It is usually assumed that the quantum jump occurs at a random instant of
time $t=s$ with a given probability density $\rho\left( s\right) >0$ on the
positive half of line $\mathbb{R}^{+}$, $\int_{0}^{\infty}\rho\left(
s\right) \mathrm{d}s=1$. If $H$ and $S$ commute, the single-point
discontinuous in $t$ stochastic evolution can formally be described by the
time-dependent Schr\"{o}dinger equation $i\hbar\partial_{t}\chi\left(
t\right) =H_{s}\left( t\right) \chi\left( t\right) $ with the singular
stochastic Hamiltonian 
\begin{equation}
H_{s}\left( t\right) =H+i\hbar\delta_{s}\left( t\right) \ln S,  \label{1.1}
\end{equation}
where $\delta_{s}\left( t\right) =\delta\left( t-s\right) =\delta _{t}\left(
s\right) $ is the Dirac $\delta$-function of $z=s-t$. Indeed, integrating
the time-dependent Hamiltonian $H_{r}\left( s\right) $ over $r$ from $0$ to $%
t$ for a fixed $s\in\mathbb{R}$, one can obtain 
\begin{equation*}
V\left( t,s\right) =e^{-\frac{i}{\hbar}\int_{0}^{t}H_{s}\left( r\right) 
\mathrm{d}r}=e^{-\frac{i}{\hbar}tH}S^{\Delta_{0}^{t}\left( s\right) }=e^{%
\frac{i}{\hbar}\left( s-t\right) H}S^{\Delta_{0}^{t}\left( s\right) }e^{-%
\frac{i}{\hbar}sH},
\end{equation*}
where $\Delta_{0}^{t}\left( s\right) =\int_{0}^{t}\delta_{r}\left( s\right) 
\mathrm{d}r$ is identified with the indicator function $1_{[0,t)}$ of the
interval $[0,t)$ for a $t>0$ (at $t\leq0$ it is zero if $s>0$). The right
hand side is the form of the unitary stochastic evolution $V\left(
t,s\right) $ which should remain valid even if the operators $H$ and $S$ do
not commute. First the evolution is conservative and continuous, $V\left(
t,s\right) =e^{-\frac{i}{\hbar}tH}$ for $t\in\lbrack0,s)$, then the quantum
jump $S$ is applied at $t=s$, and at $t>s$ the evolution is again
continuous, described by the Hamiltonian $H$. As it was noted in \cite{Bel95}%
, the rigorous form of the stochastic Schr\"{o}dinger equation which gives
such solution even for noncommuting $H$ and $S$ in the positive direction of 
$t$, is the Ito differential equation 
\begin{equation}
\mathrm{d}_{t}V\left( t,s\right) +\frac{i}{\hbar}HV\left( t,s\right) \mathrm{%
d}t=\left( S-I\right) V\left( t,s\right) \mathrm{d}1_{t}\left( s\right)
,\;t>0,\quad V\left( 0,s\right) =I.  \label{1.2}
\end{equation}
Here $\mathrm{d}_{t}V\left( t,s\right) =V\left( t+\mathrm{d}t,s\right)
-V\left( t,s\right) $ is the forward differential corresponding to an
infinitesimal increment $\mathrm{d}t>0$ at $t$, and $\mathrm{d}1_{t}\left(
s\right) =\Delta_{0}^{\mathrm{d}t}\left( s-t\right) $ is the indicator
function $\Delta_{t}^{\mathrm{d}t}\left( s\right) =1_{[t,t+\mathrm{d}%
t)}\left( s\right) $, the forward increment of the Heaviside function $%
t\mapsto1_{t}\left( s\right) =1_{0}\left( s-t\right) $, where $%
1_{0}=1_{(-\infty,0)}$. The equation (\ref{1.2}) simply means that $t\mapsto
V\left( t\right) $ for a fixed $s=z$ satisfies the usual Schr\"{o}dinger
equation $i\hbar\partial_{t}V\left( t\right) =HV\left( t\right) $ if $t\neq
s $ as $\mathrm{d}1_{t}\left( s\right) =0$ for a sufficiently small $\mathrm{%
d}t$ ($\mathrm{d}t<s-t$ if $t<s$, and any $\mathrm{d}t>0$ if $t>s$), while
it jumps, $\mathrm{d}_{t}V=\left( S-I\right) V$ at $t=s$ as $\mathrm{d}%
1_{t}\left( s\right) |_{t=s}=1\gg\mathrm{d}t.$ Integrating $\mathrm{d}%
_{z}\chi\left( z\right) =\mathrm{d}_{z}V\left( z\right) \eta$ on the domain
of the operator $H$ first from $0$ to $z=s$ with an initial condition $%
\chi\left( 0\right) =\eta$, and then from $s$ to $t$ with the initial
condition $\chi\left( s_{+}\right) :=\lim_{z\searrow s}\chi\left( z\right)
=S\chi\left( s\right) $ one can easily obtain the solution in the form $%
\chi\left( t,s\right) =V\left( t,s\right) \eta$, where 
\begin{equation}
V\left( t,z\right) =e^{-\frac{i}{\hbar}tH}S\left( z\right) ^{1_{[0,t)}\left(
z\right) },\quad S\left( z\right) =e^{\frac{i}{\hbar}zH}Se^{-\frac{i}{\hbar}%
zH}  \label{1.6}
\end{equation}
without the commutativity condition for $H$ and $S$.

Now we shall prove that the stochstic single-jump discontinuous evolution $%
V\left( t\right) $ can be treated as the interaction representation $V\left(
t\right) \chi^{0}=e^{\frac{i}{\hbar}th\left( \hat{p}\right) }\chi^{t}$ for a
deterministic strongly-continuous Schr\"{o}dinger evolution $%
\chi^{0}\mapsto\chi^{t}$ in one extra dimension $z\in\mathbb{R}$ with the
initial conditions $\chi^{0}\left( z\right) =\sqrt{\rho\left( z\right) }%
\eta\in\mathcal{H}$ localized at $z>0$: $\chi^{0}\left( z\right) =0$ at $%
z\leq0$. Here $h\left( \hat{p}\right) =-\hat{p}$ is the free Hamiltonian,
where $\hat{p}=-i\hbar\partial_{z}$ is the momentum in one extra dimension
of $z\in\mathbb{R}$.

\begin{proposition}
The described stochastic Hamiltonian problem (\ref{1.2}) is unitary
equivalent to the self-adjoint boundary-value Schr\"{o}dinger problem 
\begin{equation}
i\hbar\partial_{t}\chi^{t}\left( z\right) =\left( H+i\hbar\partial
_{z}\right) \chi^{t}\left( z\right) ,\quad\chi^{t}\left( 0_{-}\right)
=S\chi^{t}\left( 0\right)  \label{1.3}
\end{equation}
in the Hilbert space $\mathcal{H}\otimes L^{2}\left( \mathbb{R}\right) $ in
the sense that the stochastic evolution $V\left( t\right) $ at $t>0$
coincides with the unitary cocycle $V\left( t,z\right) \chi^{0}\left(
z\right) =\chi^{t}\left( z-t\right) $ resolving the boundary value problem (%
\ref{1.3}) with respect to the plane propagation $e^{t\partial_{z}}$ along $%
z $ as $\chi^{t}=e^{t\partial_{z}}V\left( t\right) \chi^{0}$, $\forall
\chi^{0}\in\mathcal{H}\otimes L^{2}\left( \mathbb{R}\right) $.
\end{proposition}

\begin{proof}
The boundary value problem (\ref{1.3}) is well defined on the space of
smooth square-integrable $\mathcal{H}$-valued functions $\chi$, and is
symmetric as $H$ is self-adjoint, and due to the unitary boundary condition 
\begin{equation*}
0=\left( \left\| \chi\left( -0\right) \right\| ^{2}-\left\| \chi\left(
0\right) \right\| ^{2}\right) =2\int\func{Re}\left\langle \chi\left(
z\right) |\chi^{\prime}\left( z\right) \right\rangle \mathrm{d}z=\frac{2}{%
\hbar}\func{Im}\left\langle \chi|\hat{h}\chi\right\rangle .
\end{equation*}
In fact, this problem is selfadjoint as it has apparently unitary solution 
\begin{equation}
\chi^{t}\left( z\right) =e^{\frac{i}{\hbar}zH}\chi_{t}\left( z+t\right)
,\quad\chi_{t}\left( s\right) =S^{\Delta_{0}^{t}\left( s\right) }\chi
_{0}\left( s\right) ,  \label{1.4}
\end{equation}
where $\chi_{0}\left( s\right) =e^{-\frac{i}{\hbar}Hz}\chi^{0}\left(
s\right) $. Indeed, substituting $\chi^{t}\left( z\right) =e^{\frac {i}{\hbar%
}zH}\chi_{0}^{t}\left( z\right) $ into the equation \{\ref{1.3}\} we obtain
the transport equation $\partial_{t}\chi_{0}^{t}\left( z\right)
=\partial_{z}\chi_{0}^{t}\left( z\right) $ with the same boundary condition $%
\chi_{0}^{t}\left( -0\right) =S\chi_{0}^{t}\left( 0\right) $ and the initial
condition $\chi_{0}^{0}=\chi_{0}$ corresponding to a $\chi^{0}\in\mathcal{H}%
\otimes L^{2}\left( \mathbb{R}\right) $. This simple initial boundary-value
problem has the obvious solution $\chi_{0}^{t}\left( z\right)
=\chi_{t}\left( z+t\right) $ with $\chi_{t}$ given in (\ref{1.4}) as 
\begin{equation}
\chi_{t}\left( s\right) =S^{1_{(0,t]}\left( t-s\right) }\chi^{0}\left(
s\right) ,\;t>0,\quad\chi_{t}\left( s\right) =S^{-1_{[-t,0)}\left( s\right)
}\chi^{0}\left( s\right) ,\;t<0.  \label{1.5}
\end{equation}

The unitarity of $S^{\Delta_{0}^{t}\left( s\right) }$ in $\mathcal{H}$ and
of shift $e^{t\partial_{z}}$ in $L^{2}\left( \mathbb{R}\right) $ implies the
unitarity of the resolving map $V^{t}:\chi^{0}\mapsto\chi^{t}$ in $\mathcal{H%
}\otimes L^{2}\left( \mathbb{R}\right) $, 
\begin{equation*}
\left\| \chi^{t}\right\| ^{2}=\left\| \chi_{t}\right\| ^{2}=\left\|
\chi_{0}\right\| ^{2}=\left\| \chi^{0}\right\| ^{2}\text{.}
\end{equation*}
Moreover, the map $t\mapsto V^{t}$ has the multiplicative representation
property $V^{r}V^{t}=V^{r+t}$ of the group $\mathbb{R}\ni r,t$ because the
map $t\mapsto S^{\Delta_{0}^{t}\left( s\right) }$ is a multiplicative
shift-cocycle, 
\begin{equation*}
S^{\Delta_{0}^{r}\left( s\right) }e^{t\partial_{s}}S^{\Delta_{0}^{t}\left(
s\right) }=e^{t\partial_{s}}S^{\Delta_{0}^{r+t}\left( s\right)
},\quad\forall r,t\in\mathbb{R}
\end{equation*}
by virtue of the additive cocycle property for the commuting $%
\Delta_{0}^{t}\left( s\right) =1_{t}\left( s\right) -1_{0}\left( s\right) $: 
\begin{equation*}
\left[ \Delta_{0}^{r}+e^{t\partial_{s}}\Delta_{0}^{t}\right] \left( s\right)
=1_{r}\left( s\right) -1_{0}\left( s\right) +1_{t}\left( s+t\right)
-1_{0}\left( s+t\right) =e^{t\partial_{s}}\Delta_{0}^{r+t}\left( s\right) .
\end{equation*}

The subtraction $\chi\left( t,z\right) =e^{\frac{i}{\hbar}t\hat{h}}\chi
^{t}\left( z\right) $ of free evolution with the generator $\hat{h}%
\chi\left( z\right) =i\hbar\partial_{z}\chi\left( z\right) $ obviously gives 
\begin{align*}
\chi\left( t,s\right) & =\chi^{t}\left( s-t\right) =e^{\frac{i}{\hbar }%
\left( s-t\right) H}\chi_{t}\left( s\right) \\
& =e^{\frac{i}{\hbar}\left( s-t\right) H}S^{\Delta_{0}^{t}\left( s\right)
}e^{-\frac{i}{\hbar}sH}\chi^{0}=V\left( t,s\right) \chi^{0}\left( s\right) ,
\end{align*}
Thus the single-point discontinuous unitary $e^{-\frac{i}{\hbar}t\hat{h}}$%
-cocycle 
\begin{equation*}
V\left( t,s\right) =e^{\frac{i}{\hbar}t\hat{h}}V^{t}=e^{\frac{i}{\hbar }%
\left( s-t\right) H}S^{\Delta_{0}^{t}\left( s\right) }e^{-\frac{i}{\hbar }%
sH},\quad t\in\mathbb{R}
\end{equation*}
with $\Delta_{0}^{t}\left( s\right) =1_{[0,t)}\left( s\right) $ for a
positive $t$ and $s\in\mathbb{R}^{+}$, solves indeed the single-jump Ito
equation (\ref{1.2}). It describes the interaction representation for the
strongly continuous unitary group evolution $V^{t}$ resolving the boundary
value problem (\ref{1.3}) with initially constant functions $\chi^{0}\left(
s\right) =\eta$ at $s>0$.
\end{proof}

\begin{remark}
The toy Schr\"{o}dinger boundary value problem (\ref{1.3}) is unphysical in
three aspects. First, the equation (\ref{1.3}) is not invariant under the
reversion of time arrow, i.e. under an isometric complex conjugation $%
\eta\mapsto\bar{\eta}$ and the reflection $t\mapsto-t$, even if $\bar {S}%
=S^{-1} $ and $\func{Im}H=0$ as the Hamiltonian $\hat{h}=i\hbar\partial_{z} $
is not real, $\func{Im}\hat{h}=\hbar\partial _{z}$. Second, a physical wave
function $\psi^{t}\left( z\right) $ should have a continuous propagation in
both directions of $z$, and at the boundary must have a jump not in the
coordinate but in momentum representation. The momentum can change its
direction but not the magnitude (conservation of momentum) in the result of
the singular interaction with the boundary. And third, the free Hamiltonian $%
\hat{h}$ must be bounded from below which is not so in the case of
hamiltonian function $h\left( z,p\right) = -p$ corresponding to the equation
(\ref{1.3}).
\end{remark}

Now we show how to rectify the first two failures of the toy model, but the
third, which is a more serious failure, will be sorted out in the next
sections by considering the toy model as a dressed limiting case.

Instead of the single wave function $\chi^{t}\left( z\right) $ on $\mathbb{R}
$ let us considering the pair $\left( \psi,\tilde{\psi}\right) $ of input
and output wave functions with 
\begin{equation*}
\psi^{t}\left( z\right) =\chi^{t}\left( z\right) ,\;z\geq0,\quad \tilde{\psi}%
^{t}\left( -z\right) =\chi^{t}\left( z\right) ,\;z<0
\end{equation*}
on the half of line $\mathbb{R}^{+}$, having the scalar product 
\begin{equation*}
\int_{0}^{\infty}\left( \left\| \psi\left( z\right) \right\| ^{2}+\left\| 
\tilde{\psi}\left( z\right) \right\| ^{2}\right) \rho\left( z\right) \mathrm{%
d}z=\int_{-\infty}^{\infty}\left\| \chi\left( z\right) \right\|
^{2}\rho\left( z\right) \mathrm{d}z\text{.}
\end{equation*}
They satisfy the system of equations 
\begin{align*}
i\hbar\partial_{t}\psi^{t}\left( z\right) & =\left( H+i\hbar\partial
_{z}\right) \psi^{t}\left( z\right) ,\quad\psi^{0}\in\mathcal{H}\otimes
L^{2}\left( \mathbb{R}^{+}\right) \\
i\hbar\partial_{t}\tilde{\psi}^{t}\left( z\right) & =\left(
H-i\hbar\partial_{z}\right) \tilde{\psi}^{t}\left( z\right) ,\quad \tilde{%
\psi}^{0}\in\mathcal{H}\otimes L^{2}\left( \mathbb{R}^{+}\right)
\end{align*}
for a quantum system interacting with a massless Dirac particle in $\mathbb{R%
}^{+}$ through the boundary condition $\tilde{\psi}^{t}\left( 0\right)
=S\psi^{t}\left( 0\right) $, where $\tilde{\psi}^{t}\left( 0\right)
=\chi^{t}\left( 0_{-}\right) $. One can show that this is indeed the
diagonal form of the Dirac equation in one dimension in the
eigen-representation of the Dirac velocity $c=-\sigma_{z}$ along $z\in%
\mathbb{R}^{+}$, with the electric and magnetic field components $u_{\pm }$,
given by the symmetric and atisymmetric parts $u\pm\tilde{u}$ of $u$ on $%
\mathbb{R}$ in the case $\func{Im}u=0$. The components of $\left( \psi,%
\tilde{\psi}\right) $ propagate independently at $z>0$ as plane waves in the
opposite directions with a spin (or polarization) oriented in the direction
of $z$, and in the scalar case $\mathcal{H}=\mathbb{C}$ are connected by the
Dirac type boundary condition $\left( 1+i\mu\right) \tilde{\psi}^{t}\left(
0\right) =\left( 1-i\mu\right) \psi^{t}\left( 0\right) $ correspondent to a
point mass $\hbar\mu$ at $z=0$. The input wave function 
\begin{equation}
\psi^{t}\left( z\right) =e^{-\frac{i}{\hbar}tH}\psi^{0}\left( z+t\right)
=e^{-\frac{i}{\hbar}t\left( \hat{h}+H\right) }\psi^{0}\left( z\right)
\label{1.7}
\end{equation}
is the solution to the equation (\ref{1.3}) at $z\in\mathbb{R}^{+}$with $%
\chi^{0}|_{z>0}=\psi^{0}$ which does not need the boundary condition at $z=0$
when solving the Cauchy problem in $t>0$. The output wave function satisfies
the reflected equation at $z>0$ and the unitary boundary condition at $z=0$: 
\begin{equation}
i\hbar\partial_{t}\tilde{\psi}^{t}\left( z\right) =\left( H-i\hbar
\partial_{z}\right) \tilde{\psi}^{t}\left( z\right) ,\quad\tilde{\psi}%
^{t}\left( 0\right) =S\psi^{t}\left( 0\right) ,  \label{1.8}
\end{equation}
It has the solution 
\begin{equation*}
\tilde{\psi}^{t}\left( z\right) =e^{-\frac{i}{\hbar}tH}\left[ \tilde{\psi }%
^{0}\left( z-t\right) 1_{t}^{\bot}\left( z\right) +S\left( t-z\right)
\psi^{0}\left( t-z\right) 1_{t}\left( z\right) \right] ,
\end{equation*}
where $1_{t}^{\bot}\left( z\right) =1-1_{t}\left( z\right) $. This can be
written in the similar way as $\psi^{t}$, 
\begin{equation}
\tilde{\psi}^{t}\left( z\right) =e^{-\frac{i}{\hbar}tH}\tilde{\psi}%
^{0}\left( z-t\right) =e^{-\frac{i}{\hbar}t\left( \check{h}+H\right) }\tilde{%
\psi}^{0}\left( z\right)  \label{1.9}
\end{equation}
with $\check{h}=-i\hbar\partial_{z}$ if $\psi^{0}\left( z\right) $ is
extended into the domain $z<0$ as 
\begin{equation}
S\left( z\right) \psi^{t}\left( z\right) =\tilde{\psi}^{t}\left( -z\right)
,\quad S\left( z\right) =e^{\frac{i}{\hbar}zH}Se^{-\frac{i}{\hbar}zH}
\label{1.10}
\end{equation}
at $t=0$. Note that reflection condition (\ref{1.10}) remains valid for all $%
t>0$ if $\psi^{t}$ is extended into the region $z<0$ by (\ref{1.7}) for all $%
t\in\mathbb{R}^{+}$: 
\begin{equation*}
\tilde{\psi}^{t}\left( -z\right) =e^{-\frac{i}{\hbar}tH}S\left( t+z\right)
\psi^{0}\left( t+z\right) =S\left( z\right) e^{-\frac{i}{\hbar}tH}\psi
^{0}\left( z+t\right) =S\left( z\right) \psi^{t}\left( z\right) .
\end{equation*}
Extending also the output wave $\tilde{\psi}^{t}$ by (\ref{1.9}) into the
region $z<0$ we obtain the continuous propagation of $\psi,\tilde{\psi}$
through the boundary in the opposite directions, with the unitary reflection
holonome connection (\ref{1.10}) for all $z\in\mathbb{R}$. If $\bar{H}=H$,
where $\bar{H}\eta=\overline{H\bar{\eta}}$ with respect to a complex
conjugation in $\mathcal{H}$, then the system of Schr\"{o}dinger equations
for the pair $\left( \psi,\tilde{\psi}\right) $ remains invariant under the
time reflection with complex conjugation up to exchange $\bar{\psi}%
^{-t}\rightleftarrows\tilde{\psi}^{t}$. Indeed, in this case the complex
conjugated hamiltonian $\overline{\hat{h}}=-i\hbar\partial_{z}$ coinsides
with the operator $\check{h}$ corresponding to $\tilde{h}\left( z,p\right)
=p=\bar {h}\left( z,p\right) $. The boundary value problem is invariant
under time reversion if $\bar{S}=S^{-1}$ as the reflection condition (\ref%
{1.10}) is extended to the negative $t$ by the exchange due to $S\left(
z\right) ^{-1}=\bar{S}\left( -z\right) .$ Thus the reversion of time arrow
is equivalent to the exchange of the input and output wave functions which
is an involute isomorphism due to 
\begin{equation*}
\int_{-\infty}^{\infty}\left\| \psi\left( z\right) \right\| ^{2}\mathrm{d}%
z=\left\| \chi\right\| ^{2}=\int_{-\infty}^{\infty}\left\| \tilde{\psi}%
\left( z\right) \right\| ^{2}\mathrm{d}z.
\end{equation*}

\section{A unitary reflection model}

As we have seen in the end of the previous section, a unitary quantum state
jump at a random instant of time $s\geq0$ is a result of solving of the toy
Schr\"{o}dinger boundary value problem in the interaction representation for
a strongly continuous unitary evolution of a Dirac particle with zero mass.
The input particle, an ''instanton'' with the state vectors defining the
input probabilities for $s=z$, has the unbounded from below kinetic energy $%
e\left( p\right) =-p$ corresponding to the constant negative velocity $%
v=e^{\prime }\left( p\right) =-1$ along the intrinsic time coordinate $z$
which does not coincide with the direction of the momentum if $p>0$. One can
interpret such strange particle as a trigger for instantaneous measurement
in a quantum system at the time $z\in\mathbb{R}^{+}$, and might like to
consider it as a normal particle, like a ''bubble'' in a cloud chamber on
the boundary of $\mathbb{R}^{d}\times\mathbb{R}^{+}$ as it was assumed in 
\cite{Bel95}, with positive kinetic energy and a non-zero mass.

Our aim is to obtain the instanton as an ultrarelativistic limit of a
quantum particle with a positive kinetic energy corresponding to a mass $%
m_{0}\geq0$. Here we shall treat the kinetic energy separately for input and
output instantons as a function of the momentum $p\in\mathbb{R}^{-}$ and $%
p\in\mathbb{R}^{+}$ respectively along a coordinate $z\in\mathbb{R}^{+}$
with the same self-adjoint operator values $e\left( p\right) \geq0$ in a
Hilbert space $\mathfrak{h}$ of its spin or other degrees of freedom. For
example one can take the relativistic mass operator-function 
\begin{equation}
e\left( p\right) =\left( p^{2}+\hbar^{2}\mu^{2}\right) ^{1/2},\quad\mu
^{2}=\mu_{0}^{2}-\triangledown^{2}  \label{2.1}
\end{equation}
in the Hilbert space $\mathfrak{h}=L^{2}\left( \mathbb{R}^{d}\right) $ which
defines the velocities $v\left( p\right) =p/e\left( p\right) =e^{\prime
}\left( p\right) $ with the same signature as $p$. At the boundary $z=0$ the
incoming particle with the negative momentum $p<0$ is reflected into the
outgoing one with the opposite momentum $-p$. The singular interaction with
the boundary causes also a quantum jump in other degrees of freedom. It is
described by the unitary operator $\sigma$ in $\mathfrak{h}$ which is
assumed to commute with $e\left( p\right) $ for each $p$ as it is in the
quantum measurement model \cite{Bel95} when $\sigma=e^{\mathbf{x}\partial_{%
\mathbf{y}}}$ with $\nabla=\partial_{\mathbf{y}}$ in \{\ref{2.1}\}.

Let $\mathfrak{h}$ be a Hilbert space with isometric complex conjugation $%
\mathfrak{h}\ni\eta\mapsto\bar{\eta}\in\mathfrak{h}$, and $L_{\mathfrak{h}%
}^{2}\left( \mathbb{R}^{-}\right) =\mathfrak{h}\otimes L^{2}\left( \mathbb{R}%
^{-}\right) $ be the space of square-integrable vector-functions $f\left(
k\right) \in\mathfrak{h}$ on the half-line $\mathbb{R}^{-}\ni k$. We denote
by $\mathcal{E}^{-}$ the isomorphic space of Fourier integrals 
\begin{equation*}
\varphi\left( z\right) =\frac{1}{2\pi}\int_{-\infty}^{0}e^{ikz}f\left(
k\right) \mathrm{d}k,\quad f\in L_{\mathfrak{h}}^{2}\left( \mathbb{R}%
^{-}\right) .
\end{equation*}
which is the Hardy class of $\mathfrak{h}$-valued functions $\varphi\in L_{%
\mathfrak{h}}^{2}\left( \mathbb{R}\right) $ having the analytical
continuation into the complex domain $\func{Im}z<0$. One can interpret $L_{%
\mathfrak{h}}^{2}\left( \mathbb{R}^{-}\right) $ as the Hilbert space of
quantum input states with negative momenta $p_{k}=\hbar k$, $k<0$ along $z\in%
\mathbb{R}$ and spin states $\eta\in\mathfrak{h}$. The generalized
eigen-functions 
\begin{equation}
\varphi_{k}\left( z\right) =\exp\left[ ikz\right] \eta_{k},\;k<0,\quad
e\left( \hbar k\right) \eta_{k}=\hbar\varepsilon_{k}\eta_{k}  \label{2.2}
\end{equation}
corresponding to spectral values $\varepsilon_{k}\in\mathbb{R}^{+}$ of $%
\varepsilon\left( k\right) =\hbar^{-1}e\left( \hbar k\right) $, are given as
the harmonic waves moving from infinity towards $z=0$ with the phase speed $%
\varsigma_{k}=\varepsilon_{k}/\left| k\right| $ along $z$. The amplitudes $%
\eta_{k}$ are arbitrary in $\mathfrak{h}$ if all $e\left( p\right) $ are
proportional to the identity operator $1$ in $\mathfrak{h}$, $%
\varepsilon\left( k\right) $ $=\varepsilon_{k}1$, as it was in the previous
section where $\mathfrak{h}=\mathcal{H}$.

The singular interaction creates the output states in the same region $z>0$
of observation where the input field is, by the momentum inversion $%
p=-p_{k}\mapsto\tilde{p}=p_{k}$, reflecting the input wave functions $%
\varphi \in\mathcal{E}^{-}$ isometrically onto 
\begin{equation*}
\tilde{\varphi}\left( s\right) =\frac{1}{2\pi}\int_{-\infty}^{0}e^{-iks}%
\tilde{f}\left( k\right) \mathrm{d}k=\sigma\varphi\left( -s\right) ,\quad
s\in\mathbb{R}_{+}
\end{equation*}
by $\tilde{f}\left( k\right) =\sigma f\left( k\right) ,$ $k<0$. The space $%
\mathcal{E}^{+}=\left\{ \tilde{\varphi}:\varphi\in\mathcal{E}^{-}\right\} $
is the conjugated Hardy subspace $\mathcal{E}^{+}=\left\{ \bar{\varphi }%
:\varphi\in\mathcal{E}^{-}\right\} $ of analytical functions $\bar{\varphi }%
\left( z\right) =\overline{\varphi\left( \bar{z}\right) }$ in $\func{Im}z>0$%
. The reflected wave function satisfies the boundary condition $\tilde{%
\varphi}\left( 0\right) =\sigma\varphi\left( 0\right) $ corresponding to the
zero probability current 
\begin{equation*}
j\left( z\right) =\left\| \tilde{\varphi}\left( z\right) \right\|
^{2}-\left\| \varphi\left( z\right) \right\| ^{2}
\end{equation*}
at $z=0$, and together with the input wave function $\varphi\left( s\right)
,s\geq0$ represents the Hilbert square norms (total probability) in $%
\mathcal{E}^{-}$ and $\mathcal{E}^{+}$ by the sum of the integrals over the
half-region $\mathbb{R}_{+}$: 
\begin{equation*}
\int_{-\infty}^{\infty}\left\| \varphi\left( z\right) \right\| ^{2}\mathrm{d}%
z=\int_{0}^{\infty}\left( \left\| \varphi\left( s\right) \right\|
^{2}+\left\| \widetilde{\varphi}\left( s\right) \right\| ^{2}\right) \mathrm{%
d}s=\int_{-\infty}^{\infty}\left\| \tilde{\varphi }\left( z\right) \right\|
^{2}\mathrm{d}z.
\end{equation*}

The free dynamics of the input and output wave functions can be described as
the unitary propagation 
\begin{align}
\varphi^{t}\left( z\right) & =\frac{1}{2\pi}\int_{-\infty}^{0}e^{ik\left(
t\varsigma\left( k\right) +z\right) }f\left( k\right) \mathrm{d}k=\left[
e^{-it\hat{\varepsilon}}\varphi\right] \left( z\right) ,  \label{2.3} \\
\widetilde{\varphi}^{t}\left( z\right) & =\frac{1}{2\pi}\int_{-%
\infty}^{0}e^{ik\left( t\varsigma\left( k\right) -z\right) }\tilde{f}\left(
k\right) \mathrm{d}k=\left[ e^{-it\check{\varepsilon}}\widetilde{\varphi }%
\right] \left( z\right) ,  \notag
\end{align}
of a superposition of the harmonoc eigen-functions (\ref{2.2}) in the
negative and positive direction of $z\in\mathbb{R}$ respectively with the
same phase speeds $\varsigma_{k}>1$ which are the eigen-values of the
positive operators $\varsigma\left( k\right) =\left| k\right|
^{-1}\varepsilon\left( k\right) $. The generating self-adjoint operators $%
\hat{\varepsilon}$, $\check{\varepsilon}$ are the restrictions $\hat{%
\varepsilon}=\varepsilon \left( i\partial_{z}\right) |\mathcal{D}^{-}$, $%
\check{\varepsilon }=\varepsilon\left( i\partial_{z}\right) |\mathcal{D}^{+} 
$ of the kinetic energy operator given by the symmetric function $%
\varepsilon\left( p\right) $ on its symmetric dense domain $\mathcal{D}%
\subseteq L_{\mathfrak{h}}^{2}\left( \mathbb{R}\right) $, to the dense
domains $\mathcal{D}^{\mp}=\mathcal{D}\cap\mathcal{E}^{\mp}$ in the
invariant subspaces $\mathcal{E}^{\mp}\subseteq L_{\mathfrak{h}}^{2}\left( 
\mathbb{R}\right) $.

Instead of dealing with the free propagation of the input-output pair $%
\left( \varphi,\tilde{\varphi}\right) $ at the region $z>0$ with the
boundary condition $\tilde{\varphi}^{t}\left( 0\right)
=\sigma\varphi^{t}\left( 0\right) $, it is convenient to introduce just one
wave function 
\begin{equation}
\phi^{t}\left( z\right) =\varphi^{t}\left( z\right) ,\;\func{Re}%
z\geq0,\quad\phi^{t}\left( z\right) =\tilde{\varphi}^{t}\left( -z\right) ,\;%
\func{Re}z<0  \label{2.5}
\end{equation}
considering the reflected wave as propagating in the negative direction into
the region $z<0$. Each $\phi\left( -z\right) $ is a Hardy class function $%
\sigma\varphi$ at $z>0$ , as well as it is Hardy class function $%
\varphi\left( -z\right) $ at $z>0$, but the continuity of the analytical
wave functions $\varphi\left( z\right) $ at $\func{Re}z=0$ corresponds to
the left discontinuity $\phi\left( 0_{-}\right) =\sigma \phi\left( 0\right) $
of 
\begin{equation*}
\phi\left( z\right) =1_{0}\left( -z\right) \varphi\left( z\right)
+1_{0}\left( z\right) \sigma\varphi\left( z\right) ,\quad1_{0}\left(
z\right) =\{_{1,\;z<0}^{0,\;z\geq0},
\end{equation*}
where $\phi\left( 0_{-}\right) $ is defined as the left lower sectorial
limit of $\phi\left( z\right) $ at $\func{Re}z\nearrow0$, $\func{Im}%
z\nearrow0$. Obviously the Hilbert subspace $\sigma^{\hat {1}_{0}}\mathcal{E}%
^{-}\subset L_{\mathfrak{h}}^{2}\left( \mathbb{R}\right) $ of such wave
functions is isomorphic to $\mathcal{E}^{-}$ by the unitary operator $%
\sigma^{\hat{1}_{0}}=I+\hat{1}_{0}\left( \sigma-I\right) $, where $\hat {1}%
_{0}$ is the multiplication operator of $\varphi\left( z\right) $ by $1$ if $%
z<0$, and by $0$ if $z\geq0.$ The unitary evolution group $\upsilon
^{t}=\sigma^{\hat{1}_{0}}e^{-it\hat{\varepsilon}}\sigma^{-\hat{1}_{0}},t\in%
\mathbb{R}$ for 
\begin{equation}
\phi^{t}\left( z\right) =\varphi^{t}\left( z\right) +1_{0}\left( z\right)
\left( \sigma-1\right) \varphi^{t}\left( z\right) =\sigma ^{1_{0}\left(
z\right) }\varphi^{t}\left( z\right) ,  \label{2.4}
\end{equation}
is unitary equivalent but different from the free propagation $e^{-it\hat {%
\varepsilon}}$ of $\varphi^{t}$ in $\mathbb{R}$. Each harmonic
eigen-function (\ref{2.2}) having the plane wave propagation 
\begin{equation*}
\varphi_{k}^{t}\left( z\right) =e^{-i\varepsilon_{k}t}\varphi_{k}\left(
z\right) =\varphi_{k}\left( z+\varsigma_{k}t\right) ,
\end{equation*}
for the negative $k\in\mathbb{R}^{-}$, \ is now truncated, $\phi_{k}\left(
z\right) =e^{ikz}\sigma^{1_{0}\left( z\right) }\eta_{k}$, and propagates in
the negative direction as 
\begin{equation*}
\phi_{k}^{t}\left( z\right) =\sigma^{1_{0}\left( z\right) }\varphi
_{k}\left( z+\varsigma_{k}t\right) =e^{-i\varepsilon_{k}t}\phi_{k}\left(
z\right) \neq\phi_{k}\left( z+\varsigma_{k}t\right) ,
\end{equation*}
keeping the truncation at $z=0$. Therefore the subtraction $\phi_{t}\left(
z\right) =e^{it\hat{\varepsilon}}\phi^{t}\left( z\right) $ of the free
propagation of $\varphi^{t}$ from $\phi^{t}$ does not return it to the
initial $\phi^{0}=\sigma^{\hat{1}_{0}}\varphi^{0}$ but to $\phi_{t}=\sigma^{%
\hat{\pi }^{t}}\varphi^{0}=\upsilon_{t}\phi_{0}$, where $\hat{\pi}^{t}=e^{it%
\hat {\varepsilon}}\hat{1}_{0}e^{-it\hat{\varepsilon}}$, $%
\upsilon_{t}=\sigma ^{\hat{\pi}^{t}}\sigma^{-\hat{1}_{0}}$, and $%
\phi_{0}=\phi^{0}$. Thus we have proved the following proposition for the
particular case $\varkappa=0$ of an operator $\varkappa\in\mathbb{R}$,
defined in the Proposition 1 of previous section as $\hbar\varkappa=H$ in $%
\mathfrak{h}=\mathcal{H}$.

Let $\varkappa $ be a selfajoint operator in $\mathfrak{h}$, and $\epsilon
_{\varkappa }\left( z\right) =e^{-i\varkappa z}$ be the correspondent
one-parameter unitary group in $\mathfrak{h}$. Below we shall denote by $%
\hat{\epsilon}_{\varkappa }$ and $\check{\epsilon}_{\varkappa }$ the
operators of pointwise multiplication by the functions $\epsilon _{\varkappa
}:z\mapsto \epsilon _{\varkappa }\left( z\right) $ and $\widetilde{\epsilon
_{\varkappa }}:z\mapsto \epsilon _{\varkappa }\left( -z\right) $ of $z\in 
\mathbb{R}$ respectively. Both these operators are unitary in the Hilbert
space $L_{\mathfrak{h}}^{2}\left( \mathbb{R}\right) $. If $\hbar \hat{\gamma}%
=\hat{h}$ is an operator in $L_{\mathfrak{h}}^{2}\left( \mathbb{R}\right) $
which is given as a pseudo-differential operator $h\left( z,\frac{\hbar }{i}%
\partial _{z}\right) =\hbar \gamma \left( z,i\partial _{z}\right) $, the
operator-function 
\begin{equation}
\gamma \left( z,\kappa +\varkappa \right) =\epsilon _{\varkappa +\kappa
}^{\ast }\left( z\right) \gamma \left( z,i\partial _{z}\right) \epsilon
_{\varkappa +\kappa }\left( z\right) \equiv \gamma _{\varkappa +\kappa
}\left( z\right) ,  \label{2.6}
\end{equation}
defines the symbol $\gamma _{\varkappa }\left( z,\kappa \right) =\gamma
\left( z,\varkappa +\kappa \right) $ of the operator 
\begin{equation*}
\hat{\gamma}_{\varkappa }=\hat{\epsilon}_{\varkappa }^{\ast }\hat{\gamma}%
\hat{\epsilon}_{\varkappa }\equiv \gamma _{\varkappa }\left( z,i\partial
_{z}\right) .
\end{equation*}
It is discribed on the exponential functions $\epsilon _{\kappa }\left(
z\right) =e^{-i\kappa z}$ as the pseudo-differential operator 
\begin{equation*}
\left[ \hat{\gamma}_{\varkappa }\epsilon _{\kappa }\eta \right] \left(
z\right) =\gamma _{\varkappa }\left( z,i\partial _{z}\right) e^{-i\kappa
z}\eta =e^{-i\kappa z}\gamma _{\varkappa }\left( z,\kappa \right) \eta
,\;\eta \in \mathfrak{h}\text{.}
\end{equation*}

\begin{proposition}
Let $\mathcal{E}_{0}^{-}=\mathcal{E}^{-}$ be the Hardy class of $L_{%
\mathfrak{h}}^{2}\left( \mathbb{R}\right) $, $\mathcal{E}_{\varkappa
}^{-}\subset L_{\mathfrak{h}}^{2}\left( \mathbb{R}\right) $ be the Hilbert
space of functions $\varphi =\hat{\epsilon}_{\varkappa }^{\ast }\varphi _{0}$
with $\varphi _{0}\in \mathcal{E}_{0}^{-}$, and $\mathcal{E}_{\varkappa
}^{+}=\check{\epsilon}_{\varkappa }^{\ast }\mathcal{E}_{0}^{+}$, where $%
\mathcal{E}_{0}^{+}=\mathcal{E}^{+}$. Let the initial boundary-value Schr%
\"{o}dinger problem 
\begin{align}
i\partial _{t}\varphi ^{t}\left( z\right) & =\varepsilon _{\varkappa }\left(
i\partial _{z}\right) \varphi ^{t}\left( z\right) ,\quad \varphi ^{0}\in 
\mathcal{E}_{\varkappa }^{-},z>0,  \label{2.7} \\
i\partial _{t}\tilde{\varphi}^{t}\left( z\right) & =\tilde{\varepsilon}%
_{\varkappa }\left( i\partial _{z}\right) \tilde{\varphi}^{t}\left( z\right)
,\quad \tilde{\varphi}^{t}\left( 0\right) =\sigma \varphi ^{t}\left(
0\right) ,  \notag
\end{align}
be defined by the generators $\hat{\varepsilon}_{\varkappa },\check{%
\varepsilon}_{\varkappa }$ given by the symbols $\varepsilon _{\varkappa
}\left( \kappa \right) =\varepsilon \left( \varkappa +\kappa \right) $, $%
\tilde{\varepsilon}_{\varkappa }\left( \kappa \right) =\varepsilon \left(
\varkappa -\kappa \right) $ respectively, where $\varepsilon \left( \kappa
\right) $ is the symmetric function of $\kappa \in \mathbb{R}$,
corresponding to the kinetic energy $e\left( p\right) =\hbar \varepsilon
\left( \hbar ^{-1}p\right) >0$. Then it is selfadjoint if the initial output
waves $\tilde{\varphi}^{0}$ are defined in $\mathcal{E}_{\varkappa }^{+}$ by 
$\tilde{\varphi}^{0}\left( -z\right) =\sigma _{\varkappa }\left( z\right)
\varphi ^{0}\left( z\right) $, $z<0$, where $\sigma _{\varkappa }=\epsilon
_{\varkappa }^{\ast }\sigma \epsilon _{\varkappa }$, by analytical
continuation of each $\varphi _{0}^{0}=\hat{\epsilon}_{\varkappa }\varphi
^{0}$ into the domain $\mathbb{R}^{-}$. The solutions to (\ref{2.7}) can be
written as 
\begin{equation}
\varphi ^{t}\left( z\right) =\phi ^{t}\left( z\right) ,z\geq 0,\;\tilde{%
\varphi}^{t}\left( -z\right) =\phi ^{t}\left( z_{-}\right) ,z\leq 0
\label{2.8}
\end{equation}
where $\phi ^{t}=e^{-it\hat{\varepsilon}_{\varkappa }}\phi _{t}$, $\phi
_{t}=\varphi ^{0}+\left( \hat{\sigma}_{\varkappa }-1\right) \hat{\pi}%
_{\varkappa }^{t}\varphi ^{0}$, $\hat{\sigma}_{\varkappa }$ is pointwise
multiplication by $\sigma _{\varkappa }\left( z\right) $, and 
\begin{equation*}
\hat{\pi}_{\varkappa }^{t}=e^{it\hat{\varepsilon}_{\varkappa }}\hat{1}%
_{0}e^{-it\hat{\varepsilon}_{\varkappa }}\equiv \pi _{\varkappa }^{t}\left(
z,i\partial _{z}\right)
\end{equation*}
is given by the symbol $\pi ^{t}\left( z,\kappa \right) $ of the
orthoprojector $\hat{\pi}^{t}=e^{it\hat{\varepsilon}}\hat{1}_{0}e^{-it\hat{%
\varepsilon}}$ as in (\ref{2.6}).
\end{proposition}

\begin{proof}
Separating the variable $t\in\mathbb{R}$ by $\varphi^{t}=e^{-i\varepsilon
_{k}t}\varphi_{k}$, $\tilde{\varphi}^{t}=e^{-i\varepsilon_{k}t}\tilde{%
\varphi }_{k}$, let us consider the stationary Schr\"{o}dinger problem 
\begin{equation}
\varepsilon_{\varkappa}\left( z,i\partial_{z}\right) \varphi_{k}\left(
z\right) =\varepsilon_{k}\varphi_{k}\left( z\right) ,\quad\tilde{\varphi }%
_{k}\left( -z\right) =\sigma_{\varkappa}\left( z\right) \varphi _{k}\left(
z\right)  \label{2.9}
\end{equation}
corresponding to the given initial and boundary conditions in (\ref{2.7}).
Here $\varphi_{k}$ is extended to the domain $\mathbb{R}^{-}$ through the
analytical continuation of $\epsilon_{\varkappa}^{\ast}\varphi_{k}$ in $%
\func{Im}z<0$, which are the generalized eigen-functions (\ref{2.2}) of $%
\hat{\varepsilon}=\varepsilon\left( i\partial_{z}\right) $ in $\mathcal{E}%
_{0}^{-}$ iff $k<0$. Due to the self-adjointness of $\hat {\varepsilon}$ in $%
\mathcal{E}^{-}$, the eigenfunctions $\varphi_{k}=\epsilon_{\varkappa+k}^{%
\ast}\eta_{k}$ of $\hat{\varepsilon}_{\varkappa}$ for (\ref{2.9}) with
negative $k$ form an orthocomplete set for the Hilbert space $\mathcal{E}%
_{\varkappa}^{-}$, and the output eigen-functions $\tilde{\varphi }%
_{k}\left( z\right) =\widetilde{\epsilon_{\varkappa+k}^{\ast}}\left(
z\right) \tilde{\eta}_{k}$, where $\tilde{\eta}_{k}=\sigma_{0}\eta_{k}$ with 
$\sigma_{0}=\rho_{0}^{-1/2}\sigma\rho_{0}^{1/2}$, form an orthocomplete set
for the Hilbert space $\mathcal{E}_{\varkappa}^{+}$. The solutions to (\ref%
{2.7}) can be written in the form (\ref{2.3}) as 
\begin{align*}
\varphi^{t}\left( z\right) & =\frac{1}{2\pi}\int_{-\infty}^{0}e^{-i%
\varepsilon\left( k\right) t}\epsilon_{\varkappa+k}^{\ast}\left( z\right)
f\left( k\right) \mathrm{d}k=\left[ e^{-it\hat{\varepsilon }%
_{\varkappa}}\varphi^{0}\right] \left( z\right) , \\
\widetilde{\varphi}^{t}\left( z\right) & =\frac{1}{2\pi}\int_{-%
\infty}^{0}e^{-i\varepsilon\left( k\right) t}\widetilde{\epsilon_{\varkappa
+k}^{\ast}}\left( z\right) \tilde{f}\left( k\right) \mathrm{d}k=\left[ e^{-it%
\check{\varepsilon}_{\varkappa}}\widetilde{\varphi}^{0}\right] \left(
z\right) ,
\end{align*}
where $\tilde{f}_{0}$ $\left( k\right) =\sigma f_{0}\left( k\right) $ are
defined as the Fourier transforms 
\begin{equation*}
f\left( k\right) =\int_{-\infty}^{\infty}\epsilon_{\varkappa+k}\left(
z\right) \varphi^{0}\left( z\right) \mathrm{d}z,\;\tilde{f}\left( k\right)
=\int_{-\infty}^{\infty}\widetilde{\epsilon_{\varkappa+k}}\left( z\right) 
\tilde{\varphi}^{0}\left( z\right) \mathrm{d}z,\;
\end{equation*}
by the initial conditions, analytically extended on the whole line $\mathbb{R%
}.$ Due to the commutativity of $\sigma$ and $\hat{\varepsilon}$ they
satisfy the connection $\tilde{\varphi}^{t}\left( -z\right)
=\sigma_{\varkappa}\left( z\right) \varphi^{t}\left( z\right) $ for all $t$,
not only for $t=0$. The time invariance of this connection and the unitarity
of the time transformation group in the Hilbert space $\mathcal{E}%
_{\varkappa}^{-}\oplus\mathcal{E}_{\varkappa}^{+}$, which follows from the
unitarity of (\ref{2.3}) in $\mathcal{E}^{\mp}\subset L_{\mathfrak{h}%
}^{2}\left( \mathbb{R}\right) $, means the self-adjointness of the problem (%
\ref{2.7}) for the pairs $\varphi^{\mp}\in L_{\mathfrak{h}}^{2}\left( 
\mathbb{R}\mathbf{,\rho}\right) $ in the domain of the generator $\hat{%
\varepsilon }_{\varkappa}\oplus\check{\varepsilon}_{\varkappa}$ with the
connection $\varphi^{+}\left( -z\right) =\sigma_{\varkappa}\left( z\right)
\varphi^{-}\left( z\right) $. Introducing 
\begin{equation*}
\phi^{t}\left( z\right) =\varphi^{t}\left( z\right) +1_{0}\left( z\right)
\left( \sigma_{\varkappa}\left( z\right) -1\right) \varphi ^{t}\left(
z\right) =\sigma_{\varkappa}\left( z\right) ^{1_{0}\left( z\right)
}\varphi^{t}\left( z\right)
\end{equation*}
as in (\ref{2.4}), and taking into account that 
\begin{equation*}
\phi^{t}\left( z_{-}\right) =\sigma_{\varkappa}\left( z\right) ^{1_{0}\left(
z_{-}\right) }\varphi^{t}\left( z\right) =\sigma_{\varkappa }\left( z\right)
^{1-1_{0}\left( -z\right) }\varphi^{t}\left( z\right)
=\sigma_{\varkappa}\left( z\right) ^{-1_{0}\left( -z\right) }\tilde{\varphi}%
^{t}\left( -z\right) ,
\end{equation*}
we obtain the representation (\ref{2.8}) as $\varphi^{t}\left( z\right) $
coincides with $\phi^{t}\left( z\right) $ at $z\geq0$ and $\tilde{\varphi }%
^{t}\left( -z\right) $ with $\tilde{\phi}^{t}\left( -z\right) =\phi
^{t}\left( z_{-}\right) $ at $z\leq0$.
\end{proof}

\begin{remark}
The Schr\"{o}dinger boundary value problem (\ref{2.7}) is physical in all
three aspects. First, the equation (\ref{2.7}) is invariant under the
reversion of time arrow, i.e. under the reflection $t\mapsto -t$ and an
isometric complex conjugation $\varphi \mapsto \bar{\varphi}$ together with
the input-output exchange $\varphi \leftrightarrows \tilde{\varphi}$ if $%
\bar{\sigma}=\sigma ^{-1}$ and $\bar{\varkappa}=\tilde{\varkappa}$, where $%
\tilde{\varkappa}\left( z\right) =\varkappa \left( -z\right) $. Second, the
wave functions $\varphi ^{t},\tilde{\varphi}^{t}$ have continuous
propagation in both directions of the momentum along $z$, and at the
boundary $z=0$ the momentum changes its direction but not the magnitude
(conservation of momentum) as the result of the boundary condition $\varphi
\left( 0\right) \mapsto \tilde{\varphi}\left( 0\right) $. And third, the
kinetic energy operator $\hat{\varepsilon}_{\varkappa }\oplus \check{%
\varepsilon}_{\varkappa }$ is bounded from below as the result of unitary
transformation of $\hat{\varepsilon}\simeq \check{\varepsilon}$. (\ref{1.3}).
\end{remark}

Indeed, from $\bar{\varepsilon}\left( \kappa\right) =\varepsilon\left(
\kappa\right) =\tilde{\varepsilon}\left( \kappa\right) $ it follows that the
symbol $\bar{\varepsilon}_{\varkappa}\left( z,\kappa\right) =\overline{%
\varepsilon_{\varkappa-\kappa}}\left( z\right) $ of the complex conjugated
operator $\overline{\varepsilon_{\varkappa}}$ is given by 
\begin{equation*}
\overline{\varepsilon_{\varkappa}}\left( z\right) =\epsilon_{-\bar {\varkappa%
}}^{\ast}\left( z\right) \varepsilon\left( i\partial_{z}\right) \epsilon_{-%
\bar{\varkappa}}\left( z\right) =\epsilon_{-\tilde{\varkappa}}^{\ast}\left(
z\right) \varepsilon\left( i\partial_{z}\right) \epsilon_{-\tilde{\varkappa}%
}\left( z\right) =\widetilde{\varepsilon _{\varkappa}}\left( z\right)
\end{equation*}
if $\bar{\varkappa}=\tilde{\varkappa}$, as $\overline{\epsilon_{\varkappa}}%
\left( z\right) =\epsilon_{-\bar{\varkappa}}\left( z\right) $ and $%
\epsilon_{-\tilde{\varkappa}}\left( z\right) =\widetilde{\epsilon
_{\varkappa}}\left( z\right) $. Therefore $\bar{\varepsilon}_{\varkappa
}\left( z,\kappa\right) =$ $\tilde{\varepsilon}_{\varkappa}\left(
z,\kappa\right) $, where $\tilde{\varepsilon}_{\varkappa}\left(
z,\kappa\right) =\widetilde{\varepsilon_{\varkappa-\kappa}}\left( z\right) $
is the symbol for the kinetic energy operator $\check{\varepsilon }%
_{\varkappa}=\widetilde{\hat{\varepsilon}_{\varkappa}}$ for the output wave $%
\tilde{\varphi}$. Thus the time reversion with complex conjugation in (\ref%
{2.7}) is equivalent to the input-output interchange $\left( \varphi ^{t},%
\tilde{\varphi}^{t}\right) \mapsto\left( \tilde{\varphi}^{t},\varphi^{t}%
\right) $ which preserves the connection between $\varphi^{t}$ and $\tilde{%
\varphi}^{t}$ as 
\begin{equation*}
\overline{\sigma_{\varkappa}}\left( z\right) =\epsilon_{-\bar{\varkappa}%
}^{\ast}\left( z\right) \overline{\sigma}\epsilon_{-\bar{\varkappa}}\left(
z\right) =\epsilon_{-\tilde{\varkappa}}^{\ast}\left( z\right) \sigma
^{-1}\epsilon_{-\tilde{\varkappa}}\left( z\right) =\widetilde{\sigma
_{\varkappa}}\left( z\right) ^{-1},
\end{equation*}
where $\widetilde{\sigma_{\varkappa}}\left( z\right) =\sigma_{\varkappa
}\left( -z\right) $ due to $\overline{\sigma}=\sigma^{-1}$.

\section{The ultrarelativistic limit.}

We shall assume here that the symmetric positive kinetic energy $e\left(
p\right) $ has the relativistic form $\left| p\right| $, or more generally, $%
e\left( p\right) =\sqrt{p^{2}+\hbar^{2}\mu^{2}}$ as it was suggested in (\ref%
{2.1}). It corresponds to the finite bounds $v_{\mp}=\mp1$ of the velocity $%
v\left( p\right) =\varepsilon^{\prime}\left( p\right) $ at $%
p\longrightarrow\mp\infty$. Note that the phase speed 
\begin{equation*}
\varsigma_{\kappa}=\varepsilon\left( \kappa\right) /\kappa=\sqrt {%
1+\varepsilon^{2}/\kappa^{2}}=\left| v\left( \hbar\kappa\right) ^{-1}\right|
,
\end{equation*}
for the momenta $p=\mp\hbar\kappa,\kappa>0$ of the harmonic eigen-waves 
\begin{equation*}
e^{-i\varepsilon_{\kappa}t}\epsilon_{\kappa}\left( z\right) =e^{-i\kappa
\left( \varsigma_{\kappa}t+z\right) },\quad e^{-i\varepsilon_{\kappa}t}%
\tilde{\epsilon}_{\kappa}\left( z\right) =e^{-i\kappa\left(
\varsigma_{\kappa}t-z\right) }
\end{equation*}
has also the limit $\varsigma=1$ at $\kappa\longrightarrow\infty$. Therefore
one should expect that the rapidly oscillating input and output waves 
\begin{equation}
\varphi^{t}\left( z\right) =e^{-i\kappa\left( t+z\right) }\psi^{t}\left(
z\right) ,\quad\tilde{\varphi}^{t}\left( z\right) =e^{-i\kappa\left(
t-z\right) }\tilde{\psi}^{t}\left( z\right) ,  \label{3.1}
\end{equation}
in the ultrarelativistic limit $p\longrightarrow\mp\infty$ will propagate as
the plane waves with 
\begin{equation}
\psi^{t}\left( z\right) =\psi\left( z+t\right) \equiv
e^{t\partial_{z}}\psi,\quad\tilde{\psi}^{t}\left( z\right) =\tilde{\psi}%
\left( z-t\right) \equiv e^{t\tilde{\partial}_{z}}\tilde{\psi}  \label{3.5}
\end{equation}
if the initial conditions are prepared in this form with slowly changing
amplitudes $\psi,\tilde{\psi}\in L_{\mathfrak{h}}^{2}\left( \mathbb{R}%
\right) $. This propagation will reproduce the boundary-reflection dynamics $%
\tilde{\psi }^{t}\left( 0\right) =\sigma\psi\left( 0\right) $ on the half
line $\mathbb{R}^{+}\ni z=s$ if the initial wave amplitudes are connected by 
$\tilde{\psi}\left( -z\right) =\sigma\psi\left( z\right) $ for all $z\in%
\mathbb{R}$. In particular, the solutions $\psi^{t}\left( s\right)
=\psi\left( s+t\right) $, $\tilde{\psi}^{t}\left( s\right) =0,t<s$, 
\begin{equation*}
\tilde{\psi}^{t}\left( s\right) =\tilde{\psi}^{t-s}\left( 0\right)
=\sigma\psi^{t-s}\left( 0\right) =\sigma\psi\left( t-s\right) ,t>s
\end{equation*}
to this Hamiltonian boundary value problem with the input wave functions 
\begin{equation*}
\psi\left( z\right) =\sqrt{\rho\left( z\right) }\eta,s>0,\quad\psi\left(
z\right) =0,z\leq0,
\end{equation*}
for the initial state-vectors $\eta\in\mathfrak{h}$ will correspond to the
single-jump stochastic dynamics in the positive direction of $t$ with
respect to the probability density $\rho>0,$ $\int_{0}^{\infty}\rho\left(
s\right) \mathrm{d}s=1$.

Below we give a precise formulation and proof of this conjecture in a more
general framework which is needed for the derivation of quantum stochastic
evolution as the boundary value problem in second quantization. But first
let us introduce the notations and illustrate this limit in this simple case.

In the following we shall use the notion of the inductive limit of an
increasing family $\left( \mathcal{E}_{\kappa}\right) _{\kappa>0}$ of
Hilbert subspaces $\mathcal{E}_{\kappa}\subseteq\mathcal{E}%
_{\kappa^{\prime}},\kappa<\kappa^{\prime}$. It is defined as the union $%
\mathcal{E}=\cup\mathcal{E}_{\kappa}$ equipped with the inductive
convergence which coinsides with the uniform convergence in one of the
subspaces $\mathcal{E}_{\kappa}$, and therefore is stronger than the
convergence in the uniform completion $\mathcal{K}=\overline{\mathcal{E}}$.
The dual inductive convergence is weaker then the convergence in $\mathcal{K}
$, and the inductive operator convergence in $\mathcal{E}$ is defined as the
operator convergence on each $\mathcal{E}_{\kappa}$ into one of $\mathcal{E}%
_{\kappa^{\prime}}\subseteq\mathcal{K}$.

Let $\mathcal{G}^{-}=\cup\mathcal{E}_{\kappa}^{-}$, $\mathcal{G}^{+}=\cup%
\mathcal{E}_{\kappa}^{+}$ be the inductive limits for the increasing family $%
\left( \mathcal{E}_{\kappa}^{-},\mathcal{E}_{\kappa}^{+}\right) _{\kappa>0}$
of Hardy classes $\mathcal{E}_{\kappa}^{-}=\hat{\epsilon}_{\kappa}^{\ast}%
\mathcal{E}^{-}\supset\mathcal{E}_{\kappa^{\circ}}^{-}$, $\mathcal{E}%
_{\kappa}^{+}=\check{\epsilon}_{\kappa}^{\ast}\mathcal{E}^{+}\supset\mathcal{%
E}_{\kappa^{\circ}}^{+}$, $\kappa^{\circ}<\kappa$ in the notations of the
previous section. Both $\mathcal{G}^{-},\mathcal{G}^{+}$ are dense in $L_{%
\mathfrak{h}}^{2}\left( \mathbb{R}\right) $, consist of the
square-intergable $\mathfrak{h}$-valued functions $\psi\in\mathcal{G}^{-}$, $%
\tilde{\psi}\in\mathcal{G}^{+}$ having zero Fourier amplitudes 
\begin{equation*}
g\left( k\right) =\int_{-\infty}^{\infty}e^{-ikz}\psi\left( z\right) \mathrm{%
d}z,\quad\tilde{g}\left( k\right) =\int_{-\infty}^{\infty}e^{ikz}\tilde{\psi}%
\left( z\right) \mathrm{d}z
\end{equation*}
for all $k\geq\kappa$ with sufficiently large $\kappa>0$. If $\psi \in%
\mathcal{E}_{\kappa}^{-}$ and $\tilde{\psi}\in\mathcal{E}_{\kappa}^{+}$,
then $\varphi=\epsilon_{\kappa}\psi\in\mathcal{E}^{-}$, $\tilde{\varphi }=%
\tilde{\epsilon}_{\kappa}\tilde{\psi}\in\mathcal{E}^{+}$, and the free
propagation (\ref{2.3}) can be written in the form (\ref{3.1}) with 
\begin{align*}
\psi^{t} & =e^{i\kappa t}\hat{\epsilon}_{\kappa}^{\ast}\varphi^{t}=\hat{%
\epsilon}_{\kappa}^{\ast}e^{-i\left( \hat{\varepsilon}-\kappa1\right) t}\hat{%
\epsilon}_{\kappa}\psi\equiv\psi_{\kappa}^{t}, \\
\tilde{\psi}^{t} & =e^{i\kappa t}\check{\epsilon}_{\kappa}^{\ast}\tilde{%
\varphi}^{t}=\check{\epsilon}_{\kappa}^{\ast}e^{-i\left( \check{\varepsilon}%
-\kappa1\right) t}\check{\epsilon}_{\kappa}\tilde{\psi }\equiv\tilde{\psi}%
_{\kappa}^{t}.
\end{align*}
These unitary transformations in $\mathcal{E}_{\kappa}^{-}$ and in $\mathcal{%
E}_{\kappa}^{+}$, written as 
\begin{equation}
\psi_{\kappa}^{t}\left( z\right) =e^{-it\omega_{\kappa}\left( i\partial
_{z}\right) }\psi\left( z\right) ,\quad\tilde{\psi}_{\kappa}^{t}\left(
z\right) =e^{-it\tilde{\omega}_{\kappa}\left( i\partial_{z}\right) }\tilde{%
\psi}\left( z\right) ,  \label{3.2}
\end{equation}
are generated by the selfadjoint operators 
\begin{align}
\omega_{\kappa}\left( i\partial_{z}\right) & =e^{i\kappa z}\left(
\varepsilon\left( i\partial_{z}\right) -\kappa\right) e^{-i\kappa
z}=\varepsilon\left( \kappa+i\partial_{z}\right) -\kappa,\;  \label{3.3} \\
\tilde{\omega}_{\kappa}\left( i\partial_{z}\right) & =e^{-i\kappa z}\left(
\varepsilon\left( i\partial_{z}\right) -\kappa\right) e^{i\kappa
z}=\varepsilon\left( \kappa-i\partial_{z}\right) -\kappa  \notag
\end{align}
They leave all subspaces $\mathcal{E}_{\kappa^{\circ}}^{-}$ and $\mathcal{E}%
_{\kappa^{\circ}}^{+}$ invariant respectively, however their generators $%
\hat{\omega}_{\kappa},\check{\omega}_{\kappa}$ are not positive definite for
a sufficiently large $\kappa$, and are not unitary equivalent for different $%
\kappa$ as 
\begin{align*}
\hat{\epsilon}_{\varkappa}\hat{\omega}_{\kappa}\hat{\epsilon}_{\varkappa
}^{\ast} & =\hat{\varepsilon}_{\kappa^{\circ}}-\kappa1=\hat{\omega}%
_{\kappa^{\circ}}-\varkappa1,\quad \\
\check{\epsilon}_{\varkappa}\check{\omega}_{\kappa}\check{\epsilon}%
_{\varkappa}^{\ast} & =\check{\varepsilon}_{\kappa^{\circ}}-\kappa 1=\check{%
\omega}_{\kappa^{\circ}}-\varkappa1,
\end{align*}
where $\varkappa=\kappa-\kappa^{\circ}$. Thus we have to prove that the
propagation (\ref{3.3}) has the inductive limit form of plane propagation (%
\ref{3.5}) at $\kappa\longrightarrow\infty$ corresponding to the Dirac form
of the limits 
\begin{equation*}
\lim_{\kappa\rightarrow\infty}\omega_{\kappa}\left( i\partial_{z}\right)
=i\partial_{z},\quad\lim_{\kappa\rightarrow\infty}\tilde{\omega}_{\kappa
}\left( i\partial_{z}\right) =-i\partial_{z}
\end{equation*}
for the Schr\"{o}dinger generators (\ref{3.3}).

Another thing which we are going to prove for obtaining the single-jump
stochastic limit is that the truncated wave 
\begin{equation*}
\chi_{\kappa}^{t}=e^{-it\hat{\omega}_{\kappa}}\chi_{\kappa,t},\quad
\chi_{\kappa,t}=\psi+\left( 1-\sigma\right) \hat{\pi}_{\kappa}^{t}\psi
\end{equation*}
representing the pair (\ref{3.3}) on the half-line $\mathbb{R}^{+}\ni z$ as
in (\ref{2.8}), has the discontinuous limit 
\begin{equation}
\chi^{t}\left( z\right) =\chi_{t}\left( z+t\right)
,\quad\chi_{t}=\psi+\left( 1-\sigma\right) \hat{1}_{t}\psi.  \label{3.6}
\end{equation}
Here $\hat{1}_{t}=e^{-it\partial_{z}}\hat{1}_{0}e^{it\partial_{z}}$ is
pointwise multiplication by the characteristic function $1_{t}$ of the
interval $-\infty<z<t$ which we shall obtain as the inductive limit of the
orthoprojector 
\begin{equation}
\hat{\pi}_{\kappa}^{t}=e^{it\varepsilon\left( \kappa+i\partial_{z}\right) }%
\hat{1}_{0}e^{-it\varepsilon\left( \kappa+i\partial_{z}\right) }=e^{it\hat{%
\omega}_{\kappa}}\hat{1}_{0}e^{-it\hat{\omega}_{\kappa}}  \label{3.4}
\end{equation}
at $\kappa\longrightarrow\infty$. This results are formulated in the
following proposition in full generality and notation of the proposition 2.

\begin{proposition}
Let $\mathcal{G}^{-}$ be the Hilbert inductive limit of Hardy classes $%
\mathcal{E}_{\kappa }^{-}=\hat{\epsilon}_{\kappa }^{\ast }\mathcal{E}^{-}$, $%
\mathcal{G}_{\varkappa }^{-}\subset L_{\mathfrak{h}}^{2}\left( \mathbb{R}%
,\rho \right) $ be the Hilbert space of functions $\psi =\hat{\epsilon}%
_{\varkappa }^{\ast }\psi _{0}$ with $\psi _{0}\in \mathcal{G}^{-}\equiv 
\mathcal{G}_{0}^{-}$, and $\mathcal{G}_{\varkappa }^{+}=\check{\epsilon}%
_{\varkappa }^{\ast }\mathcal{G}^{+}$, where $\mathcal{G}^{+}=\cup \mathcal{E%
}_{\kappa }^{+}\equiv \mathcal{G}_{0}^{+}$. Let the initial boundary-value
Schr\"{o}dinger problem 
\begin{align}
i\partial _{t}\psi _{\kappa }^{t}\left( z\right) & =\omega _{\varkappa
,\kappa }\left( i\partial _{z}\right) \psi _{\kappa }^{t}\left( z\right)
,\quad \psi _{\kappa }^{0}=\psi \in \mathcal{G}_{\varkappa }^{-},z>0,
\label{3.7} \\
i\partial _{t}\tilde{\psi}_{\kappa }^{t}\left( z\right) & =\tilde{\omega}%
_{\varkappa ,\kappa }\left( i\partial _{z}\right) \tilde{\psi}_{\kappa
}^{t}\left( z\right) ,z>0,\;\tilde{\psi}_{\kappa }^{t}\left( 0\right)
=\sigma \tilde{\psi}_{\kappa }^{t}\left( 0\right) ,  \notag
\end{align}
be defined by the generators 
\begin{equation*}
\hat{\omega}_{\varkappa ,\kappa }=\hat{\epsilon}_{\varkappa }^{\ast }\hat{%
\omega}_{\kappa }\hat{\epsilon}_{\varkappa },\quad \check{\omega}_{\varkappa
,\kappa }=\check{\epsilon}_{\varkappa }^{\ast }\check{\omega}_{\kappa }%
\check{\epsilon}_{\varkappa }
\end{equation*}
with the symbols $\omega _{\kappa },\tilde{\omega}_{\kappa }$ given in (\ref%
{3.3}), (\ref{2.1}), and the initial $\tilde{\psi}_{\kappa }^{0}=\tilde{\psi}
$ defined in $\mathcal{G}_{\varkappa }^{+}$ as $\tilde{\psi}\left( -z\right)
=\sigma _{\varkappa }\left( z\right) \psi \left( z\right) $, $z<0$ by
analytical continuation of each $\psi _{0}=\hat{\epsilon}_{\varkappa }\psi $
into the domain $\mathbb{R}^{-}$. Then the solutions to (\ref{3.7})
inductively converge to 
\begin{equation}
\psi ^{t}\left( z\right) =\chi ^{t}\left( z\right) ,z\geq 0,\;\tilde{\psi}%
^{t}\left( -z\right) =\chi ^{t}\left( z_{-}\right) ,z\leq 0  \label{3.8}
\end{equation}
where $\chi ^{t}\left( z\right) =\epsilon _{\varkappa }\left( t\right) \chi
_{t}\left( z+t\right) $, and $\;\chi _{t}=\psi +\left( \hat{\sigma}%
_{\varkappa }-1\right) \hat{1}_{t}\psi .$
\end{proposition}

\begin{proof}
First let us note that the generators in (\ref{3.7}) have the formal limits 
\begin{align*}
\lim_{\kappa\rightarrow\infty}\left[ \hat{\omega}_{\varkappa,\kappa}\psi%
\right] \left( z\right) & =\epsilon_{\varkappa}^{\ast}\left( z\right)
i\partial_{z}\left[ \epsilon_{\varkappa}{}\psi\right] \left( z\right)
=\left( \varkappa+i\partial_{z}\right) \psi\left( z\right) , \\
\lim_{\kappa\rightarrow\infty}\left[ \check{\omega}_{\varkappa,\kappa}\tilde{%
\psi}\right] \left( z\right) & =\widetilde{\epsilon_{\varkappa }^{\ast}}%
\left( z\right) i\tilde{\partial}_{z}\left[ \tilde{\epsilon }_{\varkappa}{}%
\tilde{\psi}\right] \left( z\right) =\left( \varkappa +i\tilde{\partial}%
_{z}\right) \tilde{\psi}\left( z\right)
\end{align*}
with $\tilde{\partial}_{z}=-\partial_{z}$. This follows from (\ref{3.4}) and 
$i\partial_{z}\epsilon_{\varkappa}=\varkappa\epsilon_{\varkappa}$, $%
\partial_{z}\widetilde{\epsilon_{\varkappa}}=i\tilde{\varkappa}\widetilde{%
\epsilon_{\varkappa}}$ as $\widetilde{\epsilon_{\varkappa}}\left( z\right)
=\epsilon_{-\varkappa}\left( z\right) $. Thus we have to prove that the
solutions to (\ref{3.7}) have the limits $\psi=\lim\psi_{\kappa}$, $\tilde{%
\psi}=\lim\tilde{\psi}_{\kappa}$ in $\mathcal{G}_{\varkappa}^{\mp}$
coinciding with the solutions to the Dirac boundary value problem 
\begin{align*}
i\partial_{t}\psi^{t}\left( z\right) & =\left( \varkappa+i\partial
_{z}\right) \psi^{t}\left( z\right) ,\quad\psi^{0}=\psi\in\mathcal{G}%
_{\varkappa}^{-},z>0, \\
i\partial_{t}\tilde{\psi}^{t}\left( z\right) & =\left( \varkappa +i\tilde{%
\partial}_{z}\right) \tilde{\psi}^{t}\left( z\right) ,z>0,\;\tilde{\psi}%
^{t}\left( 0\right) =\sigma_{0}\psi^{t}\left( 0\right)
\end{align*}
with the initial $\tilde{\psi}^{0}$ analytically defined as $\tilde{\psi}%
^{0}\left( -z\right) =\sigma_{\varkappa}\left( z\right) \psi^{0}\left(
z\right) $ in order to keep the solution $\tilde{\psi}^{t}$ also in $%
\mathcal{G}_{\varkappa}^{-}$ for all $t$.

Let us do this using the isomorphisms $\hat{\epsilon}_{\varkappa}$ $\check{%
\epsilon}_{\varkappa}$ of the dense subspaces $\mathcal{G}_{\varkappa
}^{\mp} $ and $\mathcal{G}_{0}^{\mp}\subset L_{\mathfrak{h}}^{2}\left( 
\mathbb{R}\right) $. Due this the boundary value problem (\ref{3.7}) is
equivalent to 
\begin{align*}
i\partial_{t}\psi_{0,\kappa}^{t}\left( z\right) & =\omega_{\kappa}\left(
i\partial_{z}\right) \psi_{0,\kappa}^{t}\left( z\right) ,\quad
\psi_{0,\kappa}^{0}=\psi_{0}\in\mathcal{G}_{0}^{-},z>0 \\
i\partial_{t}\tilde{\psi}_{0,\kappa}^{t}\left( z\right) & =\tilde{\omega }%
_{\kappa}\left( i\partial_{z}\right) \tilde{\psi}_{0,\kappa}^{t}\left(
z\right) ,z>0,\;\tilde{\psi}_{0,\kappa}^{t}\left( 0\right) =\sigma _{0}%
\tilde{\psi}_{0,\kappa}^{t}\left( 0\right) ,
\end{align*}
with $\omega_{\kappa}\left( -k\right) =\varepsilon\left( \kappa-k\right)
-\kappa=\tilde{\omega}_{\kappa}\left( k\right) $, and $\tilde{\psi }%
_{0,\kappa}^{0}\left( -z\right) =\sigma\psi_{0}\left( z\right) $ as $%
\sigma_{\kappa}=\epsilon_{\kappa}^{\ast}\sigma\epsilon_{\kappa}=\sigma$ for
any scalar $\kappa$. Thus we are to find the ultrarelativistic limit of the
solutions 
\begin{align}
\left[ e^{-it\hat{\omega}_{\kappa}}\psi_{0}\right] \left( z\right) & =\frac{1%
}{2\pi}\int_{-\infty}^{\kappa}e^{-i\left( t\omega_{\kappa}\left( -k\right)
-kz\right) }g\left( k\right) \mathrm{d}k,  \label{3.9} \\
\left[ e^{-it\check{\omega}_{\kappa}}\tilde{\psi}_{0}\right] \left( z\right)
& =\frac{1}{2\pi}\int_{-\infty}^{\kappa}e^{-i\left( t\omega_{\kappa}\left(
-k\right) +kz\right) }\tilde{g}\left( k\right) \mathrm{d}k,  \notag
\end{align}
with $\tilde{g}$ $\left( k\right) =\sigma g\left( k\right) $ at $%
\kappa\longrightarrow\infty$. Here the Fourier amplitudes 
\begin{equation*}
g\left( k\right) =\int_{-\infty}^{\infty}e^{-ikz}\psi_{0}\left( z\right) 
\mathrm{d}z,\quad\tilde{g}\left( k\right) =\int_{-\infty}^{\infty}e^{ikz}%
\tilde{\psi}_{0}\left( z\right) \mathrm{d}z,\;
\end{equation*}
are defined by analytical continuation of the initial conditions $\psi_{0}\in%
\hat{\epsilon}_{\kappa^{\circ}}^{\ast}\mathcal{E}_{0}^{-}$, $\tilde{\psi }%
_{0}\in\check{\epsilon}_{\kappa^{\circ}}^{\ast}\mathcal{E}_{0}^{+}$ for a $%
\kappa^{\circ}<\kappa$ such that the integration in (\ref{3.9}) can be
restricted to $k<\kappa^{\circ}$ due to $g\left( k\right) =0=\tilde {g}%
\left( k\right) $ for all $k\geq\kappa^{\circ}$. Therefore the proof that
the unitary evolution (\ref{3.9}) inductively converges to the plane
propagation $e^{t\partial_{z}}\psi_{0},e^{t\tilde{\partial}_{z}}\tilde{\psi }%
_{0}$ resolving this problem at $\kappa\longrightarrow\infty$ can be reduced
to finding an estimate of the integral 
\begin{equation*}
I\left( \kappa^{\circ},\kappa\right) =\frac{1}{2\pi}\int_{-\infty}^{\kappa^{%
\circ}}\left\| \left( e^{-i\left( k+\omega_{\kappa}\left( -k\right) \right)
t}-1\right) g\left( k\right) \right\| ^{2}\mathrm{d}k.
\end{equation*}
It gives the value to the mean square distances 
\begin{equation*}
\left\| e^{-t\partial_{z}}\psi_{0,\kappa}^{t}-\psi_{0}\right\| ^{2}=I\left(
\kappa^{\circ},\kappa\right) =\left\| e^{-t\tilde{\partial}_{z}}\tilde{\psi }%
_{0,\kappa}^{t}-\tilde{\psi}_{0}\right\| ^{2}
\end{equation*}
of $\psi_{0,\kappa}^{t}\left( z-t\right) $ from $\psi_{0}\in\hat{\epsilon }%
_{\kappa^{\circ}}^{\ast}\mathcal{E}_{0}^{-}$ and of $\tilde{\psi}_{0,\kappa
}^{t}\left( z+t\right) $ from $\tilde{\psi}_{0}\in\check{\epsilon}%
_{\kappa^{\circ}}^{\ast}\mathcal{E}_{0}^{+}$.

To this end we shall use the inequality 
\begin{equation*}
\left( \varkappa^{2}+\mu^{2}\right) ^{1/2}-\varkappa<\frac{1}{2}\frac {%
\mu^{2}}{\varkappa},\quad\forall\varkappa>\left| \mu\right|
\end{equation*}
for the monotonously increasing function $k+\omega_{\kappa}\left( -k\right)
<\kappa^{\circ}+\omega_{\kappa}\left( -\kappa^{\circ}\right) $ of $%
k<\kappa^{\circ}$. We shall treat separately the three cases in (\ref{2.1}):
the scalar massless case $\mu_{0}=0$ when $\varepsilon\left( k\right)
=\left| k\right| $, the boundedness case $\left| \mu\right| \leq m$ when $%
\varepsilon\left( k\right) \leq\sqrt{k^{2}+m^{2}}$ as in the scalar case
with $\mu=\mu_{0}>0$, and the general vector case when $\varepsilon\left(
k\right) =\left( k^{2}+\mu_{0}^{2}-\nabla^{2}\right) ^{1/2}$.

In the first case $k+\omega_{\kappa}\left( -k\right) =k-\kappa+\left|
\kappa-k\right| =0$ for all $\kappa\geq0$ and $k<\kappa$. Thus the plane
wave propagation 
\begin{equation*}
\psi_{0,\kappa}^{t}\left( z\right) =\psi_{0}\left( z+t\right) ,\quad \tilde{%
\psi}_{0,\kappa}^{t}\left( z\right) =\tilde{\psi}_{0}\left( z-t\right)
\end{equation*}
is extended by ultrarelativistic limit $\kappa\longrightarrow\infty$ from
the orthogonal Hardy classes $\mathcal{E}_{0}^{\mp}$ onto the inductive
spaces $\mathcal{G}_{0}^{\mp}$. By continuity they are uniquely defined as
the opposite plane propagations on the whole Hilbert space $L_{\mathfrak{h}%
}^{2}\left( \mathbb{R}\right) $ where they satisfy the connection $\tilde{%
\psi}_{0}\left( -z\right) =\sigma\psi_{0}\left( z\right) $.

In the second case $k+\omega_{\kappa}\left( -k\right) \leq m^{2}/2\varkappa$
for all$\quad\varkappa=\kappa-\kappa^{\circ}>\left| \mu\right| $ and $%
k<\kappa^{\circ}$. Using the inequality $\left| e^{x}-1\right| <2\left|
x\right| $ for any $x\in\mathbb{C}$ with $\left| x\right| \leq1$ we obtain
the estimate 
\begin{equation*}
\left\| I\left( \kappa^{\circ},\kappa\right) \right\| \leq\left\|
e^{-i\left( k+\omega_{\kappa}\left( -k\right) \right) t}-1\right\| <2\left|
t\right| \left\| k+\omega_{\kappa}\left( -k\right) \right\| <\left| t\right| 
\frac{m^{2}}{\varkappa}
\end{equation*}
for the integral $I\left( \kappa^{\circ},\kappa\right) $ with $\left\|
g\right\| ^{2}=\frac{1}{2\pi}\int\left\| g\left( k\right) \right\| ^{2}%
\mathrm{d}k\leq1$. Hence for any $\kappa^{\circ}>0$, $\varepsilon>0$ and
each $t\in\mathbb{R}$ there exists a $\kappa^{\prime}<\infty$ such that $%
\left\| I\left( \kappa^{\circ},\kappa\right) \right\| <\varepsilon$ for all $%
\kappa>\kappa^{\prime}$. Namely, one can take $\kappa^{\prime}=\kappa^{%
\circ}+\max\left\{ m,\left| t\right| m^{2}/\varepsilon\right\} $ such that $%
\varkappa=\kappa-\kappa^{\circ}>\kappa^{\prime}-\kappa>m$ and $\left|
t\right| m^{2}/\varkappa<\varepsilon$. Thus the plane wave propagation is
indeed the ultrarelativistic limit of (\ref{3.9}) in the inductive uniform
sense.

In the third case one should replace $\mathfrak{h}=L^{2}\left( \mathbb{R}%
^{d}\right) $ by the inductive limit $\mathfrak{h}^{\circ}=\cup\mathfrak{h}_{%
\mathbf{\kappa}}$ of Hilbert subspaces $\mathfrak{h}_{\mathbf{\kappa}}$ of
functions in $L^{2}\left( \mathbb{R}^{d}\right) $ having the localized
Fourier amplitudes $h\left( \mathbf{k}\right) =0$, $\mathbf{k}\notin\left( -%
\mathbf{\kappa},\mathbf{\kappa}\right) $ for a $\mathbf{\kappa\in}\mathbb{R}%
^{d}$. Then $\mu_{0}^{2}-\nabla^{2}<\mu_{0}^{2}+\mathbf{\kappa}^{2}$ in each 
$\mathfrak{h}_{\mathbf{\kappa}}$, and $\left\| I\left( \kappa^{\circ
},\kappa\right) \right\| <\left| t\right| \left( \mu_{0}^{2}+\mathbf{\kappa}%
^{2}\right) /\varkappa$ if $\left\| g\right\| \leq1$ for the Fourier
amplitudes of $\psi_{0}\in\mathfrak{h}_{\mathbf{\kappa}}\otimes\mathcal{E}%
_{\kappa^{\circ}}^{-}$ and of $\tilde{\psi}_{0}\in \mathfrak{h}_{\mathbf{%
\kappa}}\otimes\mathcal{E}_{\kappa^{\circ}}^{+}$, where $\mathcal{E}%
_{\kappa}^{\mp}$ are Hardy classes in $L^{2}\left( \mathbb{R}\right) $.
Hence for any $\kappa^{\circ}>0$, $\mathbf{\kappa\in}\mathbb{R}^{d}$, $%
\varepsilon>0$ and each $t\in\mathbb{R}$ there exists a $\kappa^{\prime}<%
\infty$ such that $\left\| I\left( \kappa^{\circ},\kappa\right) \right\|
<\varepsilon$ for all $\kappa>\kappa^{\prime}$, namely 
\begin{equation*}
\kappa^{\prime}=\kappa^{\circ}+\max\left\{ \sqrt{\mu_{0}^{2}+\mathbf{\kappa }%
^{2}},\left| t\right| \left( \mu_{0}^{2}+\mathbf{\kappa}^{2}\right)
/\varepsilon\right\} .
\end{equation*}
However the estimate $\left| t\right| \left( \mu_{0}^{2}+\mathbf{\kappa }%
^{2}\right) /\left( \kappa-\kappa^{\circ}\right) $ depends now on $\mathbf{%
\kappa}$ defining the choice of $g\left( k\right) $ in $\mathfrak{h}^{\circ}$
for each $k<\kappa^{\circ}$. This proves that the plane wave propagation is
the ultrarelativistic limit of (\ref{3.9}) also in the general vector case,
although not in the uniform but in the strong inductive convergence sense.

Thus the boundary value problem (\ref{3.7}) in the ultrarelativistic limit
is unitary equivalent to the plane propagations (\ref{3.5}) of opposite
waves $\psi_{0},\tilde{\psi}_{0}$ with the connection $\tilde{\psi}%
_{0}\left( -z\right) =\sigma\psi_{0}\left( z\right) $ for all $z\in\mathbb{R}
$. In the half space $z\in\mathbb{R}^{+}$ this obviously can be written as 
\begin{equation*}
\psi_{0}^{t}\left( z\right) =\chi_{0}^{t}\left( z\right) ,z\geq 0,\quad\text{
}\tilde{\psi}_{0}^{t}\left( -z\right) =\chi_{0}^{t}\left( z_{-}\right)
,z\leq0,
\end{equation*}
where $\chi_{0}^{t}\left( z\right) =\chi_{0,t}\left( z+t\right) $ is the
truncated input wave (\ref{3.6}) with $\psi_{0}$ in the capacity of $\psi$.
Returning back to $\psi^{t}=\hat{\epsilon}_{\varkappa}^{\ast}\psi_{0}^{t}$
and $\tilde{\psi}^{t}=\check{\epsilon}_{\varkappa}^{\ast}\tilde{\psi}%
_{0}^{t} $ we shall obtain the representation (\ref{3.8}) with 
\begin{equation*}
\chi^{t}\left( z\right) =\epsilon_{\varkappa}^{\ast}\left( z\right)
e^{t\partial_{z}}\epsilon_{\varkappa}\left( z\right) \chi_{t}\left( z\right)
=\epsilon_{\varkappa}\left( t\right) \chi_{t}\left( z+t\right) ,
\end{equation*}
due to continuity of $\epsilon_{\varkappa}\left( t\right) $, where $\chi
_{t}=\hat{\epsilon}_{\varkappa}^{\ast}\chi_{0,t}$ is given in (\ref{3.8}).
\end{proof}

\begin{remark}
The truncated wave $\chi ^{t}=\hat{1}_{0}^{\bot }\psi ^{t}+\hat{1}_{0}\tilde{%
\psi}^{t}$ in the interaction representation $\chi \left( t\right) =e^{i\hat{%
\gamma}t}\chi ^{t}$ with respect to the shift group generated by $\hat{\gamma%
}=i\partial _{z}$ satisfies the stochastic single-jump equation 
\begin{equation}
\mathrm{d}\chi \left( t,z\right) +i\varkappa \chi \left( t,z\right) \mathrm{d%
}t=\left( \sigma -1\right) \chi \left( t,z\right) \mathrm{d}1_{t}\left(
z\right) ,\;t>0.  \label{3.10}
\end{equation}
\end{remark}

Indeed, the dynamical group $\epsilon _{\varkappa }\left( t\right)
=e^{-i\varkappa t}$ is unitary in $\mathfrak{h}$. The one-parametric group $%
e^{t\partial _{z}}$ is apparently generated by the self-adjoint operator $%
\gamma \left( i\partial _{z}\right) =i\partial _{z}$ in $L^{2}\left( \mathbb{%
R}\right) $ which is the symbol of the generator $\hat{\gamma}$ for the
shift group evolution $e^{-it\hat{\gamma}}$. It is a unitary group in $L_{%
\mathfrak{h}}^{2}\left( \mathbb{R}\right) $ due to the shift-invariance of
the Lebesgue measure on $\mathbb{R}$. Hence the truncated wave in the
interaction representation is given by 
\begin{align*}
\chi \left( t,z\right) & =\chi ^{t}\left( z-t\right) =\epsilon _{\varkappa
}\left( t\right) \chi _{t}\left( z\right) \\
& =\epsilon _{\varkappa }^{\ast }\left( z-t\right) \epsilon _{\varkappa
}\left( z\right) \chi _{t}\left( z\right) =e^{i\left( z-t\right) \varkappa
}\chi _{0,t}\left( z\right) ,
\end{align*}
where $\chi _{0,t}=\chi _{0}+\left( \sigma -1\right) \left(
1_{t}-1_{0}\right) \chi _{0}$ with $\chi _{0}=\hat{1}_{0}^{\bot }\psi _{0}+%
\hat{1}_{0}\sigma \psi _{0}$. Taking into account that $\mathrm{d}t\mathrm{d}%
1_{t}\left( z\right) =0$ in the Hilbert space sense as it is zero almost
everywhere due to $\mathrm{d}1_{t}\left( z\right) =1\gg \mathrm{d}t\neq 0$
only for the single point $z=t$ having zero measure, we obtain 
\begin{align*}
\mathrm{d}\chi \left( t,z\right) & =e^{i\left( z-t\right) \varkappa }\left[
\left( \sigma -1\right) \mathrm{d}1_{t}\left( z\right) \chi _{0}\left(
z\right) -i\varkappa \chi _{0,t}\left( z\right) \mathrm{d}t\right] \\
& =\left[ \left( \sigma -1\right) \mathrm{d}1_{t}\left( z\right) -i\varkappa 
\mathrm{d}t\right] e^{i\left( z-t\right) \varkappa }\chi _{0,t}\left(
z\right) \\
& =\left[ \left( \sigma -1\right) \mathrm{d}1_{t}\left( z\right) -i\varkappa 
\mathrm{d}t\right] \chi \left( t,z\right) .
\end{align*}
Here we used that $\mathrm{d}1_{t}\left( z\right) =\mathrm{d}1_{0}\left(
z-t\right) =0$ if $z\neq t$ such that 
\begin{equation*}
\mathrm{d}1_{t}\left( z\right) e^{i\left( z-t\right) \varkappa }\chi
_{0,t}\left( z\right) =\mathrm{d}1_{t}\left( z\right) \chi _{0,t}\left(
z\right) =\mathrm{d}1_{t}\left( z\right) \chi _{0,z}\left( z\right)
\end{equation*}
due to $\chi _{0,t}\left( z\right) |_{t=z}=\chi _{0}\left( z\right) $ as $%
1_{t}\left( z\right) -1_{0}\left( z\right) =0$ for any $z\geq t\geq 0$. Thus
we have proved that $\chi \left( t,z\right) $ indeed satisfies the
stochastic single jump equation (\ref{3.10}) in the Hilbert space $L_{%
\mathfrak{h}}^{2}\left( \mathbb{R},\rho \right) $ of the initial conditions $%
\chi =\hat{1}_{0}^{\bot }\psi +\hat{1}_{0}\tilde{\psi}$ with respect to the
unitary group evolution $e^{t\partial _{z}}$.

Returning to the notations $\varkappa=\hbar^{-1}H$, $\sigma=S$ of the Sec. 1
in the Hilbert space $\mathfrak{h}=\mathcal{H}$ we obtain the stochastic
equation (\ref{1.2}) for the unitary cocycle $V\left( t,s\right)
=e^{-t\partial_{z}}V^{t}$, where $V^{t}=S^{\hat{1}_{0}}e^{t\left(
\partial_{z}-i\hbar ^{-1}H\right) }S^{-\hat{1}_{0}}$, as a
quantum-mechanical stochastic approximation. Namely, the toy Hamiltonian
model for the interpretation of discontinuous stochastic evolution in terms
of the strongly continuous unitary group resolving the Dirac boundary value
problem in extra dimension, is indeed the ultrarelativistic inductive limit
of a Schr\"{o}dinger boundary-value problem with bounded from below
Hamiltonian $H_{\kappa}\left( p\right) =\hbar\omega_{\kappa,\varkappa}\left(
-\hbar^{-1}p\right) $.

\end{document}